\documentclass[aps,pra,onecolumn,preprint,preprintnumbers,groupaddress,amsmath,amssymb,usenatbib,
nofootinbib,showkeys,showpacs,altaffilletter,11pt]{revtex4}
\usepackage[dvips]{graphicx}
\usepackage{dcolumn}
\usepackage{bm}
\usepackage{latexsym}
\usepackage{amsfonts}
\usepackage{subfigure}
\usepackage{amssymb}
\usepackage{amsmath}
\usepackage[colorlinks]{hyperref}
\usepackage{color}

\begin{document}

\title{Logarithmic entropy corrected holographic dark energy with $F(R,T)$ gravity}

\author{Ali R. Amani}
\email{a.r.amani@iauamol.ac.ir}
\affiliation{\centerline{Faculty of Sciences, Department of Physics, Ayatollah Amoli Branch, Islamic Azad University,}\\ P.O. Box 678, Amol, Mazandaran, Iran}

\author{A. Samiee-Nouri}
\email{abdollah\_samiee64@yahoo.com}
\affiliation{\centerline{Department of Physics, Central Tehran Branch, Islamic Azad University, Tehran, Iran.}}

\date{\today}

\begin{abstract}
In this paper, we have considered $F(R,T)$ gravity as a linear function of the curvature and torsion scalars and interacted it
with logarithmic entropy corrected holographic dark energy to evaluate cosmology solutions. The model has been investigated by FRW metric, and then the energy density and the pressure of dark energy have calculated. Also we obtained equation of state (EoS) parameter of dark energy and has plotted it with respect to both variable of cosmic time and $e$-folding number. Finally, we have described the scenario in three status: early, late and future time by $e$-folding number.
\end{abstract}
\maketitle

\section{Introduction}\label{s1}
Nowadays we know that Universe is accelerating, and this acceleration has been discovered in type Ia supernova \cite{RIES,PERS,BANA}, associated with large scale structure  \cite{TEGM,ABAK,POPA} and cosmic microwave background  \cite{BENC,SPED}. It is noted that the accelerated expansion of Universe arises form an energy, the so-called dark energy, and consists of about $70 \%$ the total energy of Universe. Also this discovery demonstrates that geometry of the Universe is very close to flat space \cite{GOMG}. Many models have been introduced to describe dark energy scenario, so that the Universe dominates with a perfect fluid by a negative pressure and an equation of state (EoS) parameter which is less than $-1$, the so-called phantom phase. Some of these models are such as the cosmological constant \cite{WEIS,PEEP}, the scalar fields \cite{Zlatev_1999, Kamenshchik_2001, Caldwell_2002, Amani_2011, Amani_2009,Amani1_2009}, the vector field \cite{Chiba_2008, Sadeghi_2010}, holographic \cite{AMAN, WEIH} and interacting model \cite{Amani_2013, AMAN1}.

In the last years, several other approaches were used to describe the accelerated expansion of the Universe. One of these theories is modified gravity theories that from the physical point of view, this prevents the complexities of the previous models especially the complicated computation of numerical solutions. Another strength of the modified gravity theories is consistent with recent observations for late accelerating Universe and dark energy. Some of these modified gravity theories are $F(R)$ gravity (arbitrary function of the Ricci scalar curvature $R$) \cite{Nojiri_2006, Nojiri_2007}, $F(T)$ gravity (arbitrary function of the torsion scalar $T$) \cite{Linder_2010, Myrzakulov_2011}, $F(G)$ gravity (arbitrary function of the Gauss-Bonnet term $G$) \cite{Nojiri_2005, Li_2007}. It should be noted that these modified gravity models establish to replace functions $F(R)$, $F(T)$ and $F(G)$ instead of gravity term $R$ in standard action. This means that modified gravitational theories are a generalization of general relativity. We also note that $F(T)$ gravity is a generalized version of teleparallel gravity originally proposed by Einstein \cite{Einstein_1928}, he had tried to redefine the unification of gravity and electromagnetism on the mathematical structure of distant parallelism by introducing of a tetrad or vierbein field, the so-called teleparallelism. Thus, instead of using the Levi-Civita connection in the framework of general relativity, we use the Weitzenb\"{o}ck connection in teleparallelism. In that case, the four-dimensional space-time manifold must be a parallelizable manifold \cite{Weitzenbock_1923,BEFE}.

In this paper, we are going to explain the late time accelerated expansion of the Universe by a combined model of $F(R)$ and $F(T)$ as $F(R,T)$ which one is functional of curvature and torsion scalars \cite{MYRR1}. Afterwards, we will interact the $F(R,T)$ gravity model with logarithmic entropy corrected holographic as a source of dark energy. Here we need to explain that holographic principle the first proposed in Ref. \cite{Hooft_1993}, and  later on this issue was used to interpret string theory \cite{SUSL}. Thereinafter, topic holographic has been introduced as an interesting candidate of dark energy. Holographic dark energy plays an important role in the black hole entropy in loop quantum gravity, so that they be derived from thermal equilibrium fluctuation, quantum fluctuation, or mass and charge fluctuations \cite{PASC, PASC1, BANE, KARA, ROVE, GHOS, CAIY}. In the present job, we intend to investigate the interacting model between $F(R,T)$ gravity and logarithmic entropy corrected holographic dark energy with this motivation that  we can describe the accelerated expansion of the Universe.

The paper is organized as follows:\\
In Sec. \ref{s2}, we review $F(R,T)$ gravity model and obtain the corresponding Friedmann equations. In Sec. \ref{s3}, we introduce the basic setup of the logarithmic entropy corrected holographic dark energy, and then will obtain the energy density and pressure of logarithmic entropy corrected holographic. In Sec. \ref{s4}, the interacting between  $F(R,T)$ gravity and logarithmic entropy corrected holographic dark energy are presented. Finally, a short summary is given in Sec. \ref{s5}.


\section{Fundamental of $F(R,T)$ gravity}\label{s2}
We start $F(R,T)$ gravity model with  the natural units  $16 \pi G = \hbar = c = 1$ by below action \cite{MYRR1}
\begin{equation}\label{S}
S = \int d^{4}x \sqrt{-g} \left(F(R,T) +L_m \right),
\end{equation}
where $g$ and $L_m$ are metric determinant and the matter Lagrangian respectively, and $F(R,T)$ is an arbitrary function of curvature scalar $R$ and torsion scalar $T$. \\
Here the curvature scalar and torsion scalar are introduced by
\begin{subequations}\label{rs1}
\begin{eqnarray}
  R &=& u+R_s,\label{rs1-1} \\
  T &=& v+T_s,\label{rs1-2}
\end{eqnarray}
\end{subequations}
where $R_s = g^{\mu \nu} R_{\mu \nu}$ (which $g^{\mu \nu}$ and $R_{\mu \nu}$ are metric and Ricci tensors) and $T_s = S_\rho^{\,\,\,\,\mu \nu} T^{\rho}_{\,\,\,\,\mu \nu}$ are the standard forms of curvature and torsion scalars, and, $u=u(a,\dot{a})$ and $v=v(a,\dot{a})$ be defined as the two arbitrary functions in terms of $a$ and $\dot{a}$, which these are scale factor and its derivative with respect to cosmic time respectively. We note that Eqs. \eqref{rs1} are a corrected choice with respect to the its standard forms, i.e., $R_s$ and $T_s$  (for more details see \cite{MYRR1}). In order to introduce the aforesaid tensors, we need to express a locally inertial frame instead of the coordinate frame. For this purpose, we consider a tetrad or vierbein field $e_i(x^\mu)$ with index $i$ running from $0$ to $3$, which one is an orthonormal basis for the tangent space at each point $x^\mu$ of the parallelizable manifold. Therefore, we can relate vierbein field to the metric as $g_{\mu \nu} = \eta_{i j} e^i_{\,\,\mu} e^j_{\,\,\nu}$ in which $ \eta_{i j}=diag(+1,-1,-1,-1)$ \cite{BEFE}. The tensor $S_\rho^{\,\,\,\,\mu \nu}$ and the torsion tensor $T^\rho_{\,\,\,\,\mu \nu}$ be witten by
\begin{subequations}\label{ts1}
\begin{eqnarray}
T^\rho{}_{\mu\nu}&=&e_A^\rho\,(\partial_\mu e^A_\nu-\partial_\nu
e^A_\mu),\label{ts1-1}\\
S_\rho{}^{\mu\nu}&=&\frac{1}{2}(K^{\mu\nu}{}_\rho+\delta^\mu_\rho
T^{\alpha\nu}{}_\alpha-\delta^\nu_\rho T^{\alpha\mu}{}_\alpha),\label{ts1-2}
\end{eqnarray}
\end{subequations}
and the contortion tensor $K^{\mu\nu}{}_\rho$ is
\begin{equation}\label{k1}
K^{\mu\nu}{}_\rho=-\frac{1}{2}(T^{\mu\nu}{}_\rho-T^{\nu\mu}{}_\rho
-T_\rho{}^{\mu\nu}).
\end{equation}
Now we use in this job the flat FRW metric in the following form
\begin{equation}\label{frw}
  ds^2=-dt^2+a^2(t)(dx^2+dy^2+dz^2),
\end{equation}
in that case, we can calculate the curvature and torsion scalar as
\begin{subequations}
\begin{eqnarray}
 R &=& u+6\left(\dot{H}+2H^2\right), \\
  T &=& v-6H^2,
\end{eqnarray}
\end{subequations}
where $H=\frac{\dot{a}}{a}$ is the Hubble parameter.\\
In order to obtain the Friedmann equations, we must find the corresponding Lagrangian of the action \eqref{S}. Therefore,  by using  Mc Lauren  expansion of phrase $F(R,T)$, and by taking conditions as $F_{RR}=F_{TT}=F_{RT}=0$ (for simplicity, $F(R,T)$ is selected as linear in terms of $R$ and $T$), we can obtain the Lagrangian in the form
\begin{equation}\label{lag1}
  L=a^3(F-TF_T-RF_R+vF_T+uF_R)-6(F_R+F_T)a\dot{a}^2+L_m,
\end{equation}
where indices denote derivative with respect to $R$ and $T$.\\
In this paper, we consider a simple particular model of $F(R,T)$ gravity as
\begin{equation}\label{frt}
F(R,T)=\mu R+\nu T,
\end{equation}
where $\mu$ and $\nu$ are constants.
\begin{subequations}\label{rho1}
\begin{eqnarray}
&\rho_{tot}=3(\mu+\nu)H^2-\frac{1}{2}(\mu\, \dot{a}\, u_{\dot{a}}+\nu \,\dot{a}\, v_{\dot{a}}-\mu\, u-\nu\, v),\label{rho1-1}\\
&-p_{tot}=(\mu+\nu)(2\dot{H}+3H^2)-\frac{1}{2}(\mu\, \dot{a}\, u_{\dot{a}}+\nu\, \dot{a}\, v_{\dot{a}}-\mu\, u-\nu\, v)-\frac{1}{6} \mu\, a (\dot{u}_{\dot{a}}-u_a)\\ \nonumber
&-\frac{1}{6} \nu\, a\, (\dot{v}_{\dot{a}}- v_a),\label{rho1-2}
\end{eqnarray}
\end{subequations}
where indices $a$ and $\dot{a}$ are derivatives with respect to themselves, and, $\rho_{tot}$ and $p_{tot}$ are total energy density and total pressure of the Universe dominated with a perfect fluid, respectively. The continuity equation of the model becomes
\begin{equation}\label{cont}
  \dot{\rho}_{tot}+3 H (\rho_{tot}+p_{tot})=0.
\end{equation}
Now in order to solve the model, we choose functions $u$ and $v$ as a power form in terms of the scale factor by
\begin{subequations}\label{uv1}
\begin{eqnarray}
u &=& \alpha \, a^n,\label{uv1-1} \\
v &=& \beta \, a^m,\label{uv1-2}
\end{eqnarray}
\end{subequations}
to substitute Eqs. \eqref{uv1} into Eqs. \eqref{rho1}, the Friedmann equations are rewritten by
\begin{subequations}\label{rho2}
\begin{eqnarray}
&\rho_{tot}=3(\mu+\nu)H^2+\frac{1}{2}\left(\mu\,\alpha\,a^n+\nu \,\beta\, a^m\right),\label{rho2-1}\\
&-p_{tot}=(\mu+\nu)\left(2\dot{H}+3H^2\right)+\frac{1}{6} \mu\,\alpha(n+3)\, a^n+\frac{1}{6} \nu\,\beta (m+3)\, a^m.\label{rho2-2}
\end{eqnarray}
\end{subequations}

 \section{Logarithmic entropy corrected holographic dark energy}\label{s3}
As we know, holographic dark energy is a candidate for description the dark energy in modern cosmology, where one has been derived from the holographic principle \cite{SUSL, THOO, PASC, PASC1}. On the one hand, derivation of holographic dark energy describes black hole that its holographic energy density depends on the entropy-area relationship $S\sim A \sim L^2$ in gravity, where $A \sim L^2$ represents the area of the horizon. On the other hand, from the point of view the quantum gravity, this interpretation can be achieved from quantum effects and creates the entropy-area relationship \cite{BANE}. Then, we can write corrected entropy-area relation as \cite{KARA, AMAN}
\begin{equation}\label{entropy}
S=\frac{A}{4}+\tilde\gamma \ln\Big(\frac{A}{4}\Big)+\tilde\beta,
\end{equation}
where $\tilde\gamma$ and $\tilde\beta$ are dimensionless constants. The corrections are due to thermal equilibrium  and quantum fluctuations \cite{ROVE, GHOS}. The second term in (\ref{entropy}) appears in a model of entropy cosmology which unifies the inflation and late time acceleration \cite{CAIY}. The $\tilde{\gamma}$
 might be extremely large due to current cosmological constraint, which inevitably brought a fine tuning problem to entropy corrected models and it is desirable to determine it by observational constrain. Taking the corrected entropy-area relation (\ref{entropy}) into account, the energy density of the holography dark energy will be
modified as well. On this basis, logarithmic entropy corrected holographic dark energy be given by \cite{WEIH}
\begin{equation}\label{rholam}
 \rho_{\Lambda}=3c^2R_h^{-2}+\gamma R_h^{-4}\ln(R_h^{2})+\beta R_h^{-4},
\end{equation}
where $c$ is a constant determined by observational fit. The future event horizon $R_h$ is defined as
\begin{equation}\label{Rh}
R_h= a\int_t^\infty \frac{dt}{a}=a\int_a^\infty\frac{da}{Ha^2},
\end{equation}
which leads to results compatible with observations. The continuity equation of logarithmic entropy corrected holographic dark energy be written by
\begin{equation}\label{coneq}
\dot{\rho}_{\Lambda}+3H(\rho_{\Lambda}+p_{\Lambda})=0,
\end{equation}
the corresponding pressure clearly obtains by the continuity equation in the form
\begin{eqnarray}\label{presant}
&p_\Lambda=-3\,c^2\,R_h^{-2}-\gamma\,R_h^{-4}\,\ln(R_h^2)+\frac{4 \,\gamma}{3\,H}\,\dot{R}_h\,R_h^{-5}\,\ln(R_h^2)+\frac{4 \,\beta}{3\,H}\,\dot{R}_h\,R_h^{-5}-\beta\,R_h^{-4}+\\\nonumber
&\frac{2 \,c^2}{H}\,\dot{R}_h\,R_h^{-3}-\frac{2 \,\gamma}{3\,H}\,\dot{R}_h\,R_h^{-5},
\end{eqnarray}
where $\dot{R}_h=H\,R_h-1$ is found the Eq. \eqref{Rh}.\\
In next section we will study interacting between $F(R,T)$ gravity and logarithmic entropy corrected holographic dark energy.
\section{interacting $F(R,T)$ gravity with logarithmic entropy corrected holographic dark energy}\label{s4}
\begin{figure}[t]
\begin{center}
\includegraphics[scale=.265]{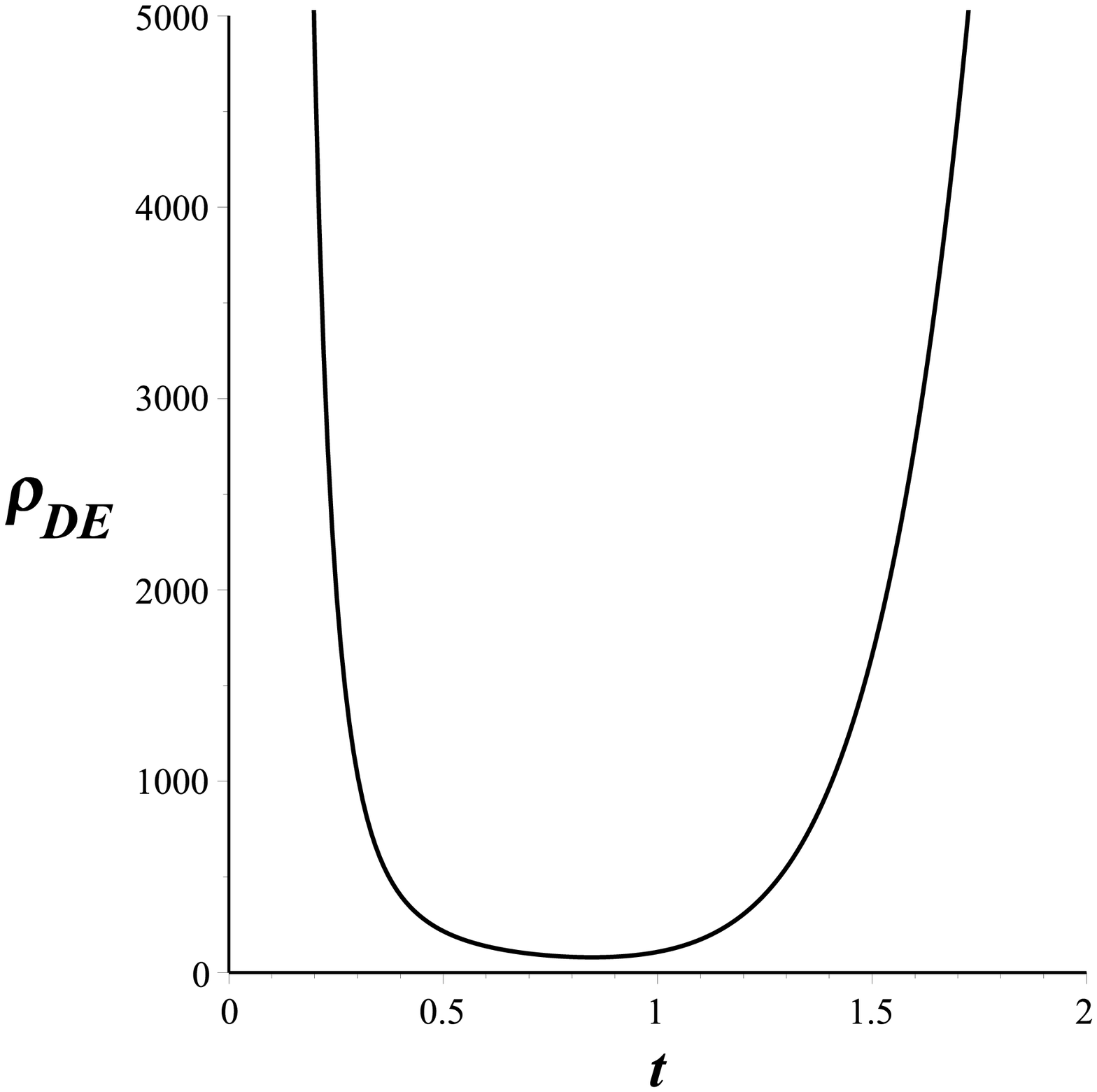}\includegraphics[scale=.265]{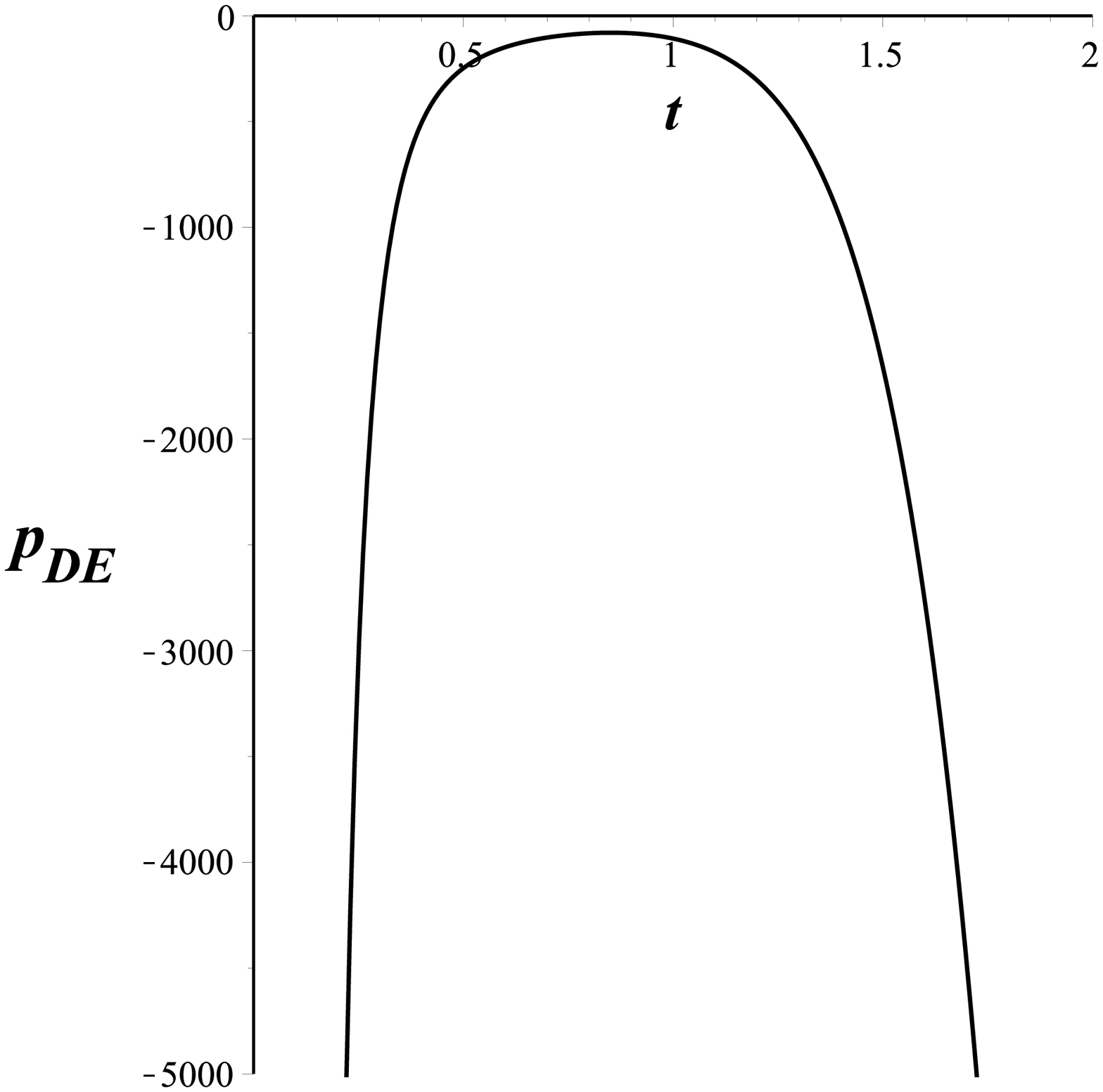}\includegraphics[scale=.265]{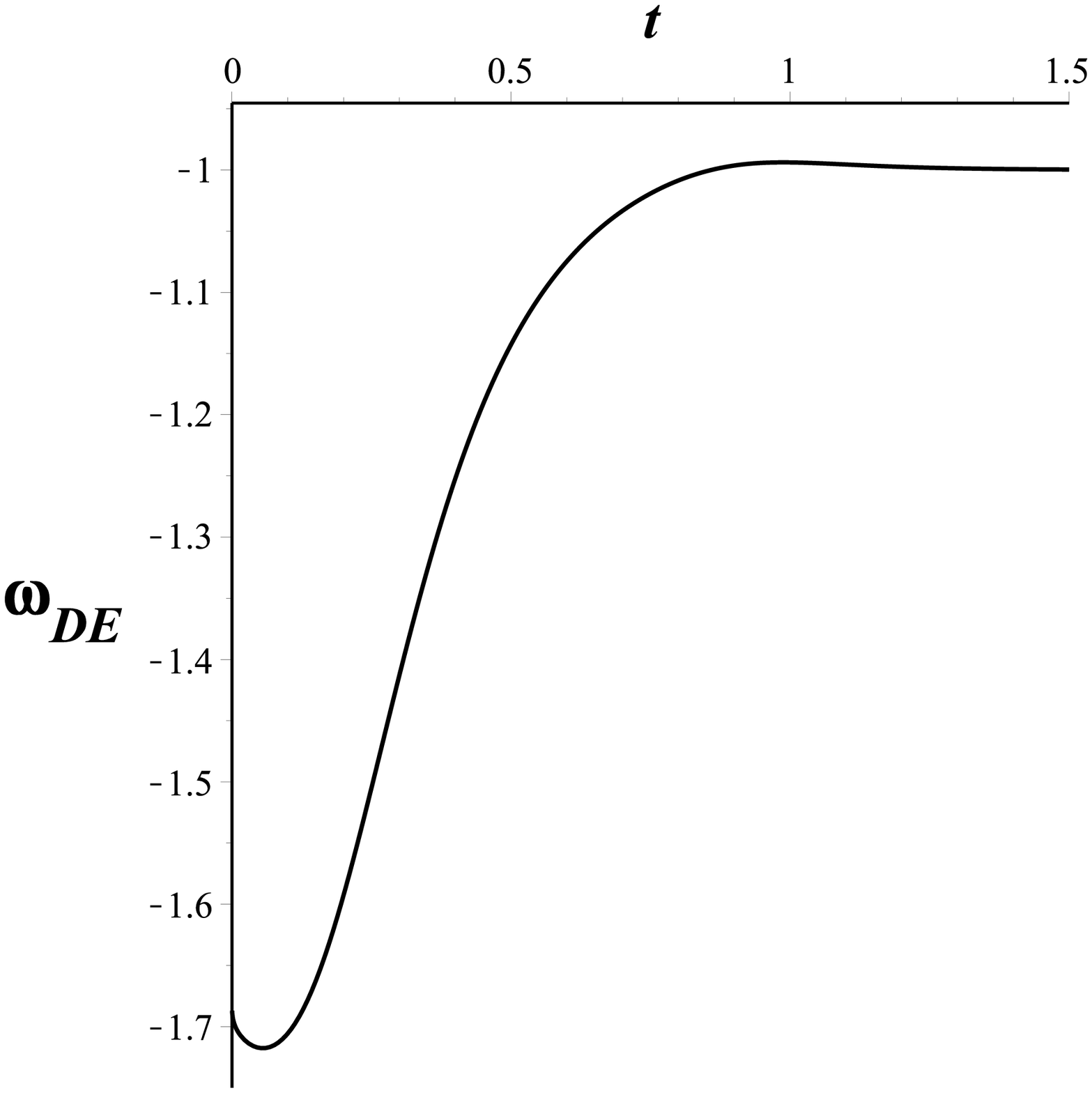}
\caption{The graphs of $\rho_{DE}$,  $p_{DE}$ and $\omega_{DE}$ in terms of cosmic time for case $s>0$ by $s = 2$, $\mu = 2$, $\nu = 2$, $c = 1$, $\gamma = 2$, $\alpha = -0.1$, $\beta = 4$, $a_0 = 2$, $m = 4$, $n = 2$ and $\lambda = 0.5$.}\label{fig1}
\end{center}
\end{figure}

\begin{figure}[t]
\begin{center}
\includegraphics[scale=.265]{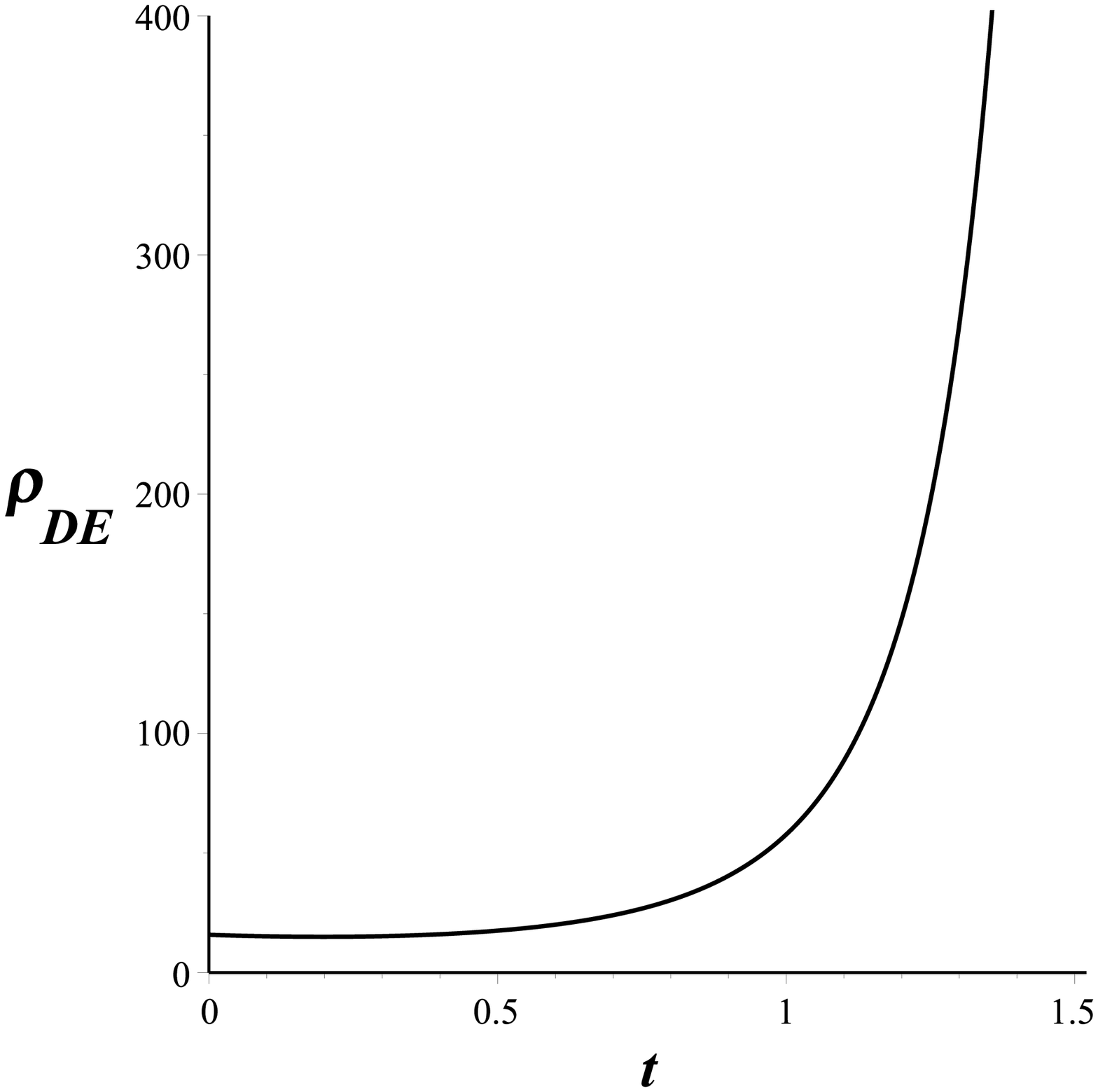}\includegraphics[scale=.265]{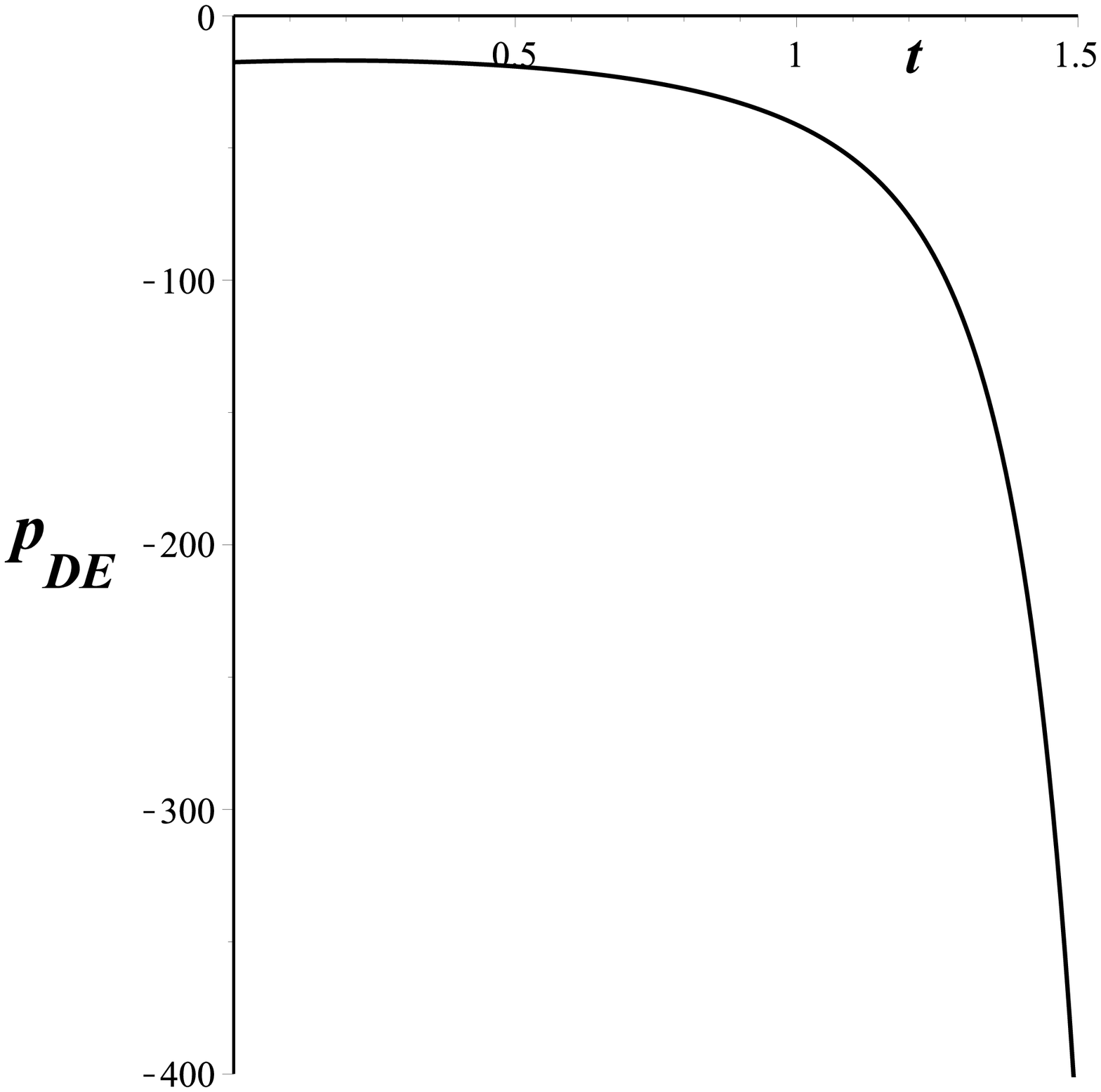}\includegraphics[scale=.265]{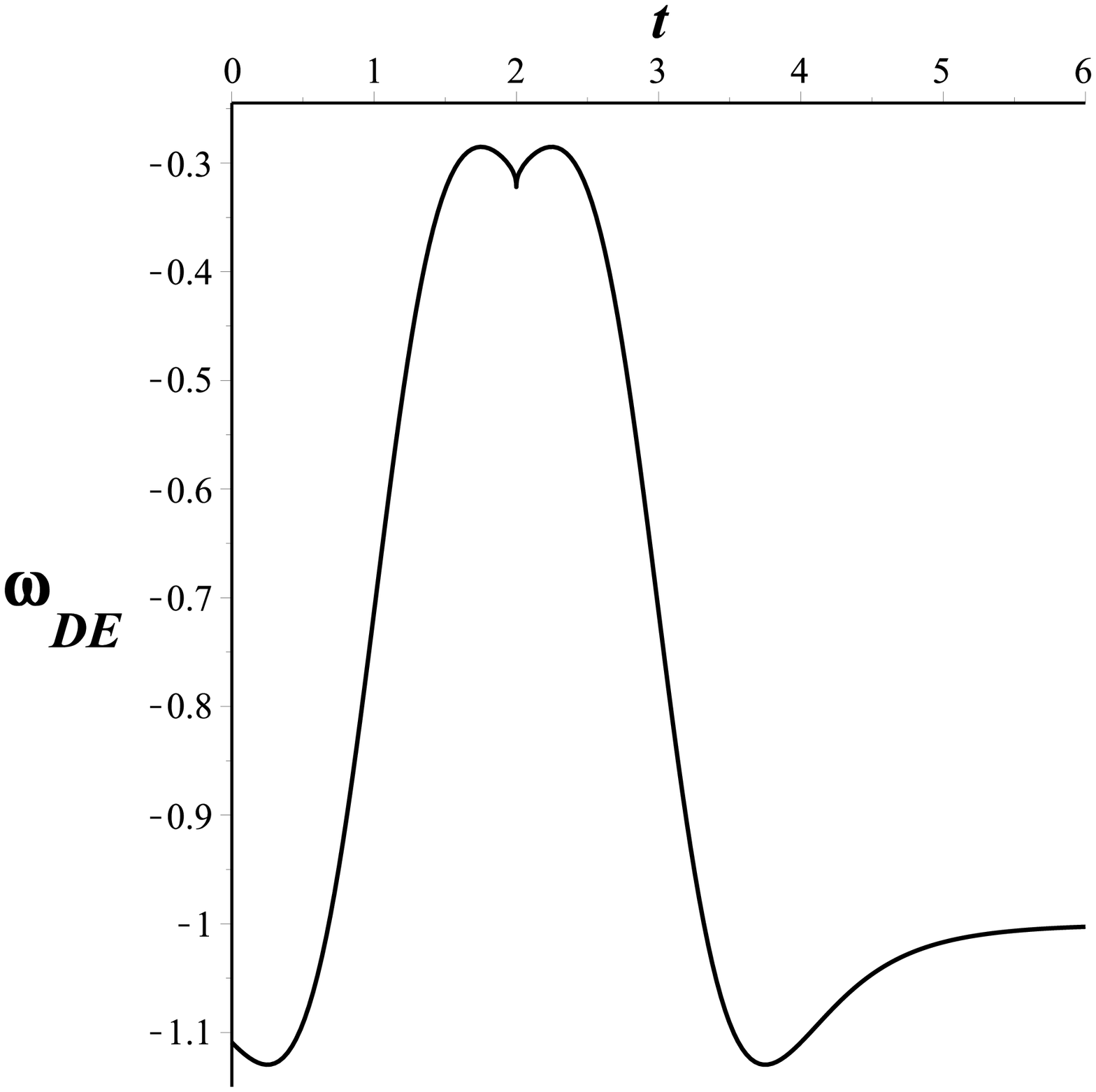}
\caption{The graphs of $\rho_{DE}$, $p_{DE}$ and $\omega_{DE}$ in terms of cosmic time for case $s<0$ by $s = -2$, $\mu = 2$, $\nu = 2$, $c= 0.5$, $\gamma = 0.5$ , $\alpha = 2$, $\beta = 2$, $a_0 = 2$, $m = -2$, $n = 2$, $\lambda = 1$ and $t_0 = 2$.}\label{fig2}
\end{center}
\end{figure}

In this section, we are going to study on the dark energy scenario by interacting between $F(R,T)$ gravity and logarithmic entropy corrected holographic dark energy in flat FRW metric. \\
Now we consider that Universe dominated with entropy corrected holographic, thus total energy density and total pressure are written as a combination of  $F(R,T)$ gravity and logarithmic entropy corrected holographic dark energy in the following form
\begin{subequations}\label{rhopres2}
\begin{eqnarray}
  \rho_{tot} &=& \rho_{DE}+\rho_{\Lambda}, \label{rhopres2-1}\\
  p_{tot} &=& p_{DE}+p_{\Lambda}. \label{rhopres2-2}
\end{eqnarray}
\end{subequations}
The continuity equation \eqref{cont} is separated as Eq. \eqref{coneq} and equation
\begin{equation}\label{coneq2}
\dot{\rho}_{DE}+3H(\rho_{DE}+p_{DE})=0,
\end{equation}
where $\rho_{DE}$ and $p_{DE}$ are found by Eqs. \eqref{rho2}, \eqref{rholam}, \eqref{presant} and \eqref{rhopres2} as
\begin{subequations}\label{rhopres3}
\begin{eqnarray}
&\rho_{DE} = 3(\mu+\nu)H^2+\frac{1}{2}\left(\mu\,\alpha\,a^n+\nu \,\beta\, a^m\right)-3c^2R_h^{-2}-\gamma R_h^{-4}\ln(R_h^{2})-\beta R_h^{-4},\label{rhopres3-1}\\
  &p_{DE} = -(\mu+\nu)\left(2\dot{H}+3H^2\right)-\frac{1}{6} \mu\,\alpha(n+3)\, a^n-\frac{1}{6} \nu\,\beta (m+3)\, a^m +\\\nonumber &3\,c^2\,R_h^{-2}+\gamma\,R_h^{-4}\,\ln(R_h^2)-\frac{4 \,\gamma}{3\,H}\,\dot{R}_h\,R_h^{-5}\,\ln(R_h^2)-\frac{4 \,\beta}{3\,H}\,\dot{R}_h\,R_h^{-5}+\beta\,R_h^{-4}-\\\nonumber
&\frac{2 \,c^2}{H}\,\dot{R}_h\,R_h^{-3}+\frac{2 \,\gamma}{3\,H}\,\dot{R}_h\,R_h^{-5}.\label{rhopres3-2}
\end{eqnarray}
\end{subequations}
The equation of state of dark energy is defined as
\begin{equation}\label{eos1}
\omega_{DE}=\frac{p_{DE}}{\rho_{DE}}.
\end{equation}
Now in order to investigate the model, we take the scale factor in the form power law as $a(t)=a_0\, t^s$, in which the scale factor increases in terms of cosmic time when $s>0$ \cite{AMAN, AMAN1}. We note that for case $s<0$, Universe is shrunk, this means that we don't have an expanding Universe. Thus, to avoid this issue we change $t$ to $-t$ so we have $a=a_0\,(-t)^s$. This choice give rise to a big rip singularity that for disappearing the singularity, we shift the origin of the cosmic time from $-t$ to $t_s-t$. Therefore, while $t$ arrives to $t_s$ occurs a big rip, this issue is solved by condition $t<t_s$. In that case, we can write the scale factor as $a=a_0\,(t_s-t)^s$ when $s<0$. Then, we can obtain the Hubble parameter for two below cases:
\begin{subequations}\label{H1}
\begin{eqnarray}
H=\frac{s}{t}\,\,\,\, \textrm{when}\,\,\,\, s>0,\label{H1-1} \\
 H=\frac{-s}{t_s-t}\,\,\,\, \textrm{when}\,\,\,\, s<0 .\label{H1-2}
\end{eqnarray}
\end{subequations}
Then $R_h$ is yielded
\begin{subequations}\label{Rh2}
\begin{eqnarray}
R_h={\frac {t}{s-1}}\,\,\,\, \textrm{when}\,\,\,\, s>0,\label{Rh2-1} \\
R_h={\frac {{t_s}-t}{-s+1}}\,\,\,\, \textrm{when}\,\,\,\, s<0 .\label{Rh2-2}
\end{eqnarray}
\end{subequations}
By inserting Eqs. \eqref{H1-1} and \eqref{Rh2-1} into Eqs. \eqref{rhopres3} we find $\rho_{DE}$ and $p_{DE}$ for case $s>0$ as
\begin{subequations}\label{rhopres4}
\begin{eqnarray}
&\rho_{DE} =3\,{\frac { \left( \mu+\nu \right) {s}^{2}}{{t}^{2}}
}+\frac{1}{2}\,\mu\,\alpha\,{a_{{0}}}^{n}{t}^{{\it ns}}+\frac{1}{2}\,\nu\,\beta\,a_{{0
}}^m{t}^{{\it ms}}-3\,{{c}^{2} \left( \frac{s-1}{t} \right) ^{2}}-\\\nonumber
&\left( \frac{s-1}{t} \right) ^{4}\left(\gamma\, \ln  \left( {\frac {t^2}{(s-1)^2}} \right) +\lambda\right),\label{rhopres4-1}\\
  &p_{DE} = \frac{1}{6st^4}\, \Big( 6\, \left( s+\frac{4}{3} \right) \gamma\, \left( s
-1 \right) ^{4}\ln  \left( {\frac {{t}^{2}}{ \left( s-1 \right) ^{2}}}
 \right) -3\,\nu\,\beta s\, \left( a_{{0}}{t}^{s} \right) ^{m}{t}^{4}-
3\,\mu\,\alpha\, \left( s \left( a_{{0}}{t}^{s} \right)  \right) ^{n}{
t}^{4}+\\\nonumber
&6\,{s}^{5}\lambda+ \left( -4\,\gamma-16\,\lambda \right) {s}^{4
}+ \left( 16\,\gamma+ \left( 18\,{c}^{2}-18\,\mu-18\,\nu \right) {t}^{
2}+4\,\lambda \right) {s}^{3}+\\\nonumber
& \left( -24\,{c}^{2}{t}^{2}-24\,\gamma+
24\,\lambda \right) {s}^{2}+ \left( -6\,{c}^{2}{t}^{2}+16\,\gamma-26\,
\lambda \right) s+12\,{c}^{2}{t}^{2}-4\,\gamma+8\,\lambda \Big).\label{rhopres4-2}
\end{eqnarray}
\end{subequations}
The graphs of $\rho_{DE}$, $p_{DE}$ and $\omega_{DE}$ are plotted for case $s>0$ in Fig. \ref{fig1}. These Figures show the variations of cosmological parameters with respect to cosmic time for the aforesaid model.  We note that chosen coefficients play the role of an important to plot the cosmological parameters. The motivation of the selections is based on that the graph $\rho_{DE}$ has the positive values, the graph $p_{DE}$ has the negative values with respect to cosmic time, and specially crossing EoS over phantom-divide-line. \\
Once more, by inserting Eqs. \eqref{H1-2} and \eqref{Rh2-2} into Eqs. \eqref{rhopres3} we find $\rho_{DE}$ and $p_{DE}$ for case $s<0$ as
\begin{subequations}\label{rhopres5}
\begin{eqnarray}
&\rho_{DE} =3\,{\frac { \left( \mu+\nu \right) {s}^{2}}{{(t_s-t)}^{2}}
}+\frac{1}{2}\,\mu\,\alpha\,{a_{{0}}}^{n}{(t_s-t)}^{{\it ns}}+\frac{1}{2}\,\nu\,\beta\,a_{{0
}}^m{(t_s-t)}^{{\it ms}}-3\,{{c}^{2} \left( \frac{-s+1}{(t_s-t)} \right) ^{2}}
-\\\nonumber
&\left( \frac{-s+1}{(t_s-t)} \right) ^{4}\left(\gamma\, \ln  \left( {\frac {(t_s-t)^2}{(-s+1)^2}} \right) +\lambda\right),\label{rhopres5-1}\\
  &p_{DE} = \frac{1}{6s(t_s-t)^4}\, \Big( 6\, \left( s+\frac{4}{3} \right) \gamma\, \left( s
-1 \right) ^{4}\ln  \left( {\frac {{(t_s-t)}^{2}}{ \left( s-1 \right) ^{2}}}
 \right) -3\,\nu\,\beta s\, \left( a_{{0}}{(t_s-t)}^{s} \right) ^{m}{(t_s-t)}^{4}-\\\nonumber
&3\,\mu\,\alpha\, \left( s \left( a_{{0}}{(t_s-t)}^{s} \right)  \right) ^{n}{
t}^{4}+6\,{s}^{5}\lambda+ \left( -4\,\gamma-16\,\lambda \right) {s}^{4
}+ \\\nonumber
&\left( 16\,\gamma+ \left( 18\,{c}^{2}-18\,\mu-18\,\nu \right) {(t_s-t)}^{
2}+4\,\lambda \right) {s}^{3}+ \left( -24\,{c}^{2}{(t_s-t)}^{2}-24\,\gamma+
24\,\lambda \right) {s}^{2}+ \\\nonumber
&\left( -6\,{c}^{2}{(t_s-t)}^{2}+16\,\gamma-26\,
\lambda \right) s+12\,{c}^{2}{(t_s-t)}^{2}-4\,\gamma+8\,\lambda \Big).\label{rhopres5-2}
\end{eqnarray}
\end{subequations}
The graphs of $\rho_{DE}$, $p_{DE}$ and $\omega_{DE}$ has been plotted for case $s<0$ in Figs. \ref{fig2}. We can see the variations of the corresponding cosmological parameters against to cosmic time by interacting $F(R,T)$ gravity with logarithmic entropy corrected holographic dark energy. The results of these Figures show us that the graphs of $\rho_{DE}$ and $p_{DE}$ are positive values and negative values against to cosmic time respectively, and the graph $\omega_{DE}$ crosses the phantom divided line. \\
\begin{figure}[h]
\begin{center}
\subfigure{\includegraphics[scale=.3]{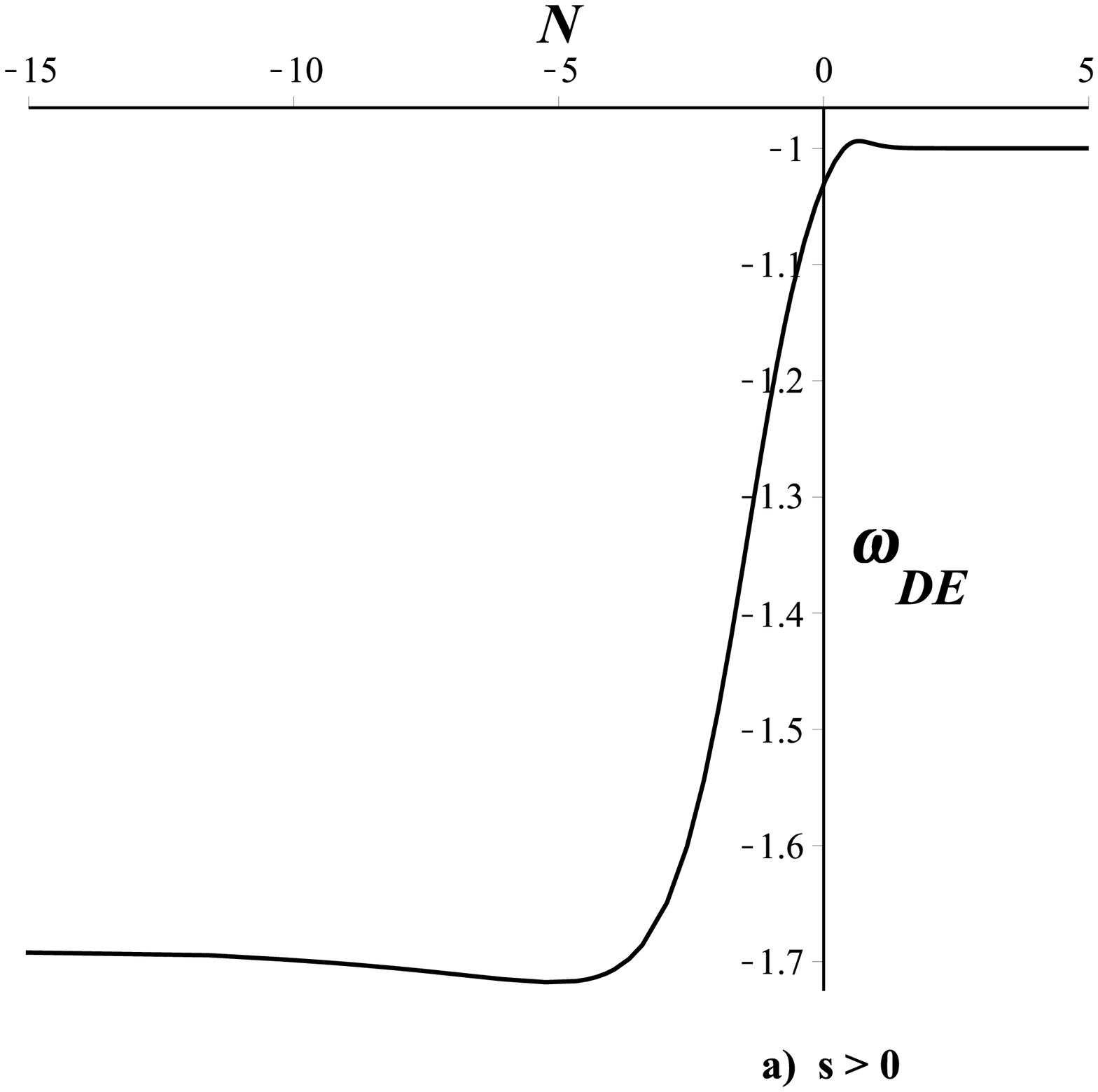}\label{fig3-1}}\qquad \qquad
\subfigure{\includegraphics[scale=.3]{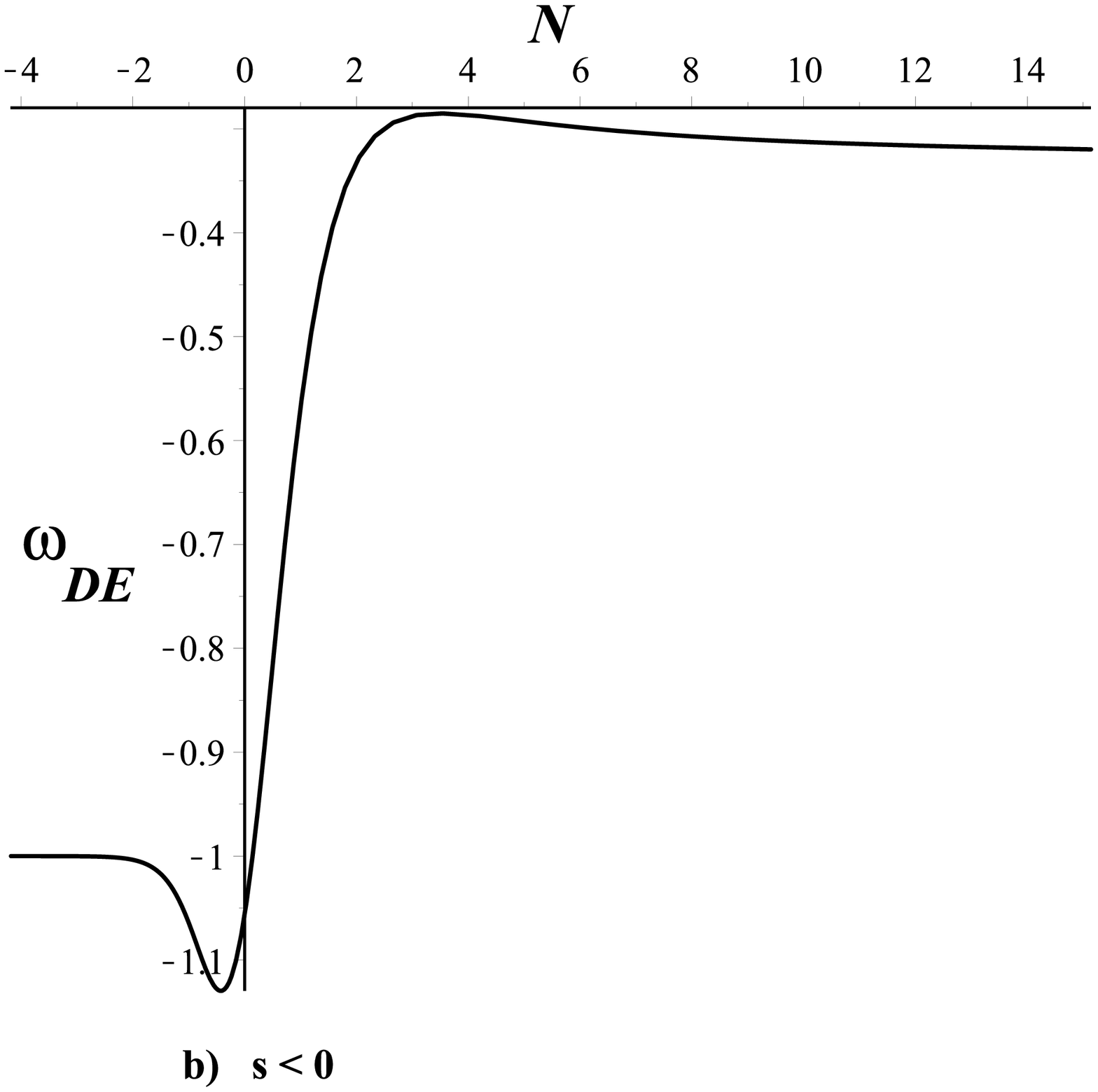}\label{fig3-2}}
\caption{The graph of $\omega_{DE}$ in terms of $N$ for both cases $s>0$ (left hand) and $s<0$ (right hand) by mentioned free parameters.}\label{fig3}
\end{center}
\end{figure}
In order to have a more complete discussion, we represent the free parameters of the model in terms of observable quantities. For this purpose, to have an accelerated Universe expansion we plot the EoS of the $F(R,T)$ gravity versus the e-folding number $N = ln(a)$. We can see the values of EoS of the current model in three cases $N \rightarrow -\infty$, late time ($N = 0$) and $N \rightarrow +\infty$ respectively with values $-1.65$, $-1.03$ and $-1$ for case $s>0$ in Fig. \ref{fig3-1}, and $-1$, $-1.055$ and $0.35$ for case $s<0$ in Fig. \ref{fig3-2}. We note that when Universe is undergoing an accelerated expansion, the EoS crosses the value of $-1$ in late time where in the scenario one can be seen in Fig. \ref{fig3}.


\section{conclusion}\label{s5}
In this paper, we have studied $F(R,T)$ gravity as an arbitrary function of curvature and torsion scalars in FRW background for describing something mysterious in the cosmology, in which for simplicity we have provided $F(R,T)$ gravity as a linear function of the curvature and torsion scalars. Afterwards, we have considered the corresponding model by interacting with logarithmic entropy corrected holographic dark energy. Should be noted that the corresponding action has been written the combination of $F(R,T)$ gravity and a matter Lagrangian. In this scenario, the friedmann equations have been obtained to take an interaction between $F(R,T)$ gravity with logarithmic entropy corrected holographic dark energy. It should be mentioned that logarithmic entropy corrected holographic dark energy has been derived from the holographic principle for description the dark energy in modern cosmology.\\
In what follows, we considered total energy density and total pressure of Universe as dominated with a perfect fluid, in which dark energy contribution is taken the combination of $F(R,T)$ gravity and logarithmic entropy corrected holographic dark energy as $\rho_{DE}=\rho_{tot}-\rho_{\Lambda}$ and $ p_{DE}=p_{tot}-p_{\Lambda}$. We note that index $tot$ is related to $F(R,T)$ gravity and one has been written in terms of functional the Hubble parameter, and index $\Lambda$ is related to logarithmic entropy corrected holographic and has been written by future event horizon.\\
 Therefore, by dividing these two function we obtained the EoS of dark energy for the scenario. Subsequently, in order to describe the scenario we have taken the scale factor as power law and by inserting it into \eqref{rhopres4} and \eqref{rhopres5}, energy density, pressure and EoS of dark energy have solved. Also we drew them with respect to cosmic time and the corresponding Figs. \ref{fig1} and \ref{fig2} showed us an accelerating Universe, because the graph of EoS crossed over phantom-divided line and these findings confirm the observational data. Finally, we tried to describe the scenario by another parameter named $e$-folding number, so we re-plotted the EoS with respect to $e$-folding number and have obtained its corresponding values in three status $N \rightarrow -\infty$, late time ($N = 0$) and $N \rightarrow +\infty$.



\begin{thebibliography}{99}

\bibitem{RIES}
Riess, A. G., et al.: Astron. J.116, 1009 (1998).

\bibitem{PERS}
Perlmutter, S.J., et al.: Astrophys. J.517, 656 (1999).

\bibitem{BANA}
Bachall, N. A., Ostriker, J.P., Perlmutter, S., Steinhardt, P.J.: Science284, 1481 (1999).

\bibitem{TEGM}
Tegmark, M., Strauss, M. A., Blanton, M. R., Abazajian, K., Dodelson, S., Sandvik, H., ... and Knapp, G. R. (2004). Cosmological parameters from SDSS and WMAP. Physical Review D, 69(10), 103501.

\bibitem{ABAK}
Abazajian, K., et al.: Astron. J.128, 502 (2004).

\bibitem{POPA}
Pope, A. C., et al.: Astrophys. J.607, 655 (2004).

\bibitem{BENC}
Bennet, C., et al.: Astrophys. J.148(Suppl.), 1 (2003).

\bibitem{SPED}
Spergel, D.N., et al.: Astrophys. J. Suppl. Ser.148, 175 (2003).

\bibitem{GOMG}
Gomero, G. I., Rebouças, M. J., and Tavakol, R. (2001). Detectability of cosmic topology in almost flat universes. Classical and Quantum Gravity, 18(21), 4461.

\bibitem{WEIS}
Weinberg, S. (1989). The cosmological constant problem. Reviews of Modern Physics, 61(1), 1.

\bibitem{PEEP}
Peebles, P. J. E., and Ratra, B. (2003). The cosmological constant and dark energy. Reviews of Modern Physics, 75(2), 559.

\bibitem{Zlatev_1999}
Zlatev, I., Wang, L., and Steinhardt, P. J. (1999). Quintessence, cosmic coincidence, and the cosmological constant. Physical Review Letters, 82(5), 896.

\bibitem{Kamenshchik_2001}
Kamenshchik, A., Moschella, U., and Pasquier, V. (2001). An alternative to quintessence. Physics Letters B, 511(2), 265-268.

\bibitem{Caldwell_2002}
Caldwell, R. R. (2002). A phantom menace? Cosmological consequences of a dark energy component with super-negative equation of state. Physics Letters B, 545(1), 23-29.

\bibitem{Amani_2011}
Amani, A. R. (2011). Stability of Quintom Model of Dark Energy in (ω, ω′) Phase Plane. International Journal of Theoretical Physics, 50(10), 3078-3088.

\bibitem{Amani_2009}
Sadeghi, J., and Amani, A. R. (2009). The solution of tachyon inflation in curved universe. International Journal of Theoretical Physics, 48(1), 14-21.

\bibitem{Amani1_2009}
Setare, M. R., Sadeghi, J., and Amani, A. R. (2009). Interacting tachyon dark energy in non-flat universe. Physics Letters B, 673(4), 241-246.

\bibitem{Chiba_2008}
Chiba, T. (2008). Initial conditions for vector inflation. Journal of Cosmology and Astroparticle Physics, 2008(08), 004.

\bibitem{Sadeghi_2010}
Sadeghi, J., Setare, M. R., Amani, A. R., and Noorbakhsh, S. M. (2010). Bouncing universe and reconstructing vector field. Physics Letters B, 685(4), 229-234.

\bibitem{AMAN}
Amani, A. R., Sadeghi, J., Farajollahi, H., and Pourali, M. (2011). Logarithmic entropy corrected holographic dark energy with nonminimal kinetic coupling. Canadian Journal of Physics, 90(1), 61-66.

\bibitem{WEIH}
Wei, H. (2009). Entropy-corrected holographic dark energy. Communications in Theoretical Physics, 52(4), 743.

\bibitem{Amani_2013}
Amani, A. R., Escamilla-Rivera, C., and Faghani, H. R. (2013) Interacting closed string tachyon with modified Chaplygin gas and its stability, Phys. Rev. D 88, 124008.

\bibitem{AMAN1}
Amani, A. R., and Pourhassan, B. (2014). Interacting closed string tachyon with generalized cosmic Chaplygin gas. International Journal of Geometric Methods in Modern Physics.

\bibitem{Nojiri_2006}
Nojiri, S. I., and Odintsov, S. D. (2006). Modified f (R) gravity consistent with realistic cosmology: From a matter dominated epoch to a dark energy universe. Physical Review D, 74(8), 086005.

\bibitem{Nojiri_2007}
Nojiri, S. I., and Odintsov, S. D. (2007). Unifying inflation with ΛCDM epoch in modified f (R) gravity consistent with Solar System tests. Physics Letters B, 657(4), 238-245.

\bibitem{Linder_2010}
Linder, E. V. (2010). Einstein’s Other Gravity and the Acceleration of the Universe. Physical Review D, 81(12), 127301.

\bibitem{Myrzakulov_2011}
Myrzakulov, R. (2011). Accelerating universe from F (T) gravity. The European Physical Journal C-Particles and Fields, 71(9), 1-8.

\bibitem{Nojiri_2005}
Nojiri, S. I., and Odintsov, S. D. (2005). Modified Gauss–Bonnet theory as gravitational alternative for dark energy. Physics Letters B, 631(1), 1-6.

\bibitem{Li_2007}
Li, B., Barrow, J. D., and Mota, D. F. (2007). Cosmology of modified Gauss-Bonnet gravity. Physical Review D, 76(4), 044027.

\bibitem{Einstein_1928}
Einstein, A. (1928). New Possibility for a Unified Field Theory of Gravitation and Electricity. Sitzungsberichte der Preussischen Akademie der Wissenschaften, Physikalisch-mathematische Klasse, 17, 224–227.

\bibitem{Weitzenbock_1923}
Weitzenb\"{o}ck, R. (1923). Invarianten theorie. P. Noordhoff.

\bibitem{BEFE}
Bengochea, G. R., and Ferraro, R. (2009). Dark torsion as the cosmic speed-up. Physical Review D, 79(12), 124019.

\bibitem{MYRR1}
Myrzakulov, R. (2012). Dark energy in F (R, T) gravity. arXiv preprint arXiv:1205.5266.

\bibitem{Hooft_1993}
't Hooft, G. (1993). Dimensional reduction in quantum gravity. Arxiv preprint gr-qc/9310026.

\bibitem{SUSL}
Susskind, L. (1995). The world as a hologram. Journal of Mathematical Physics, 36(11), 6377-6396.

\bibitem{THOO}
't Hooft, G.  (2006). The black hole horizon as a dynamical system. International Journal of Modern Physics D, 15(10), 1587-1602.

\bibitem{PASC}
Pasqua, A., Chattopadhyay, S., and Khomenko, I. (2013). A reconstruction of modified holographic Ricci dark energy in f (R, T) gravity. Canadian Journal of Physics, 91(8), 632-638.

\bibitem{PASC1}
Pasqua, A., and Khomenko, I. (2013). Interacting Ricci logarithmic entropy-corrected holographic dark energy in Brans-Dicke cosmology. International Journal of Theoretical Physics, 52(11), 3981-3993.

\bibitem{BANE}
Banerjee, R., Gangopadhyay, S., and Modak, S. K. (2010). Voros product, noncommutative Schwarzschild black hole and corrected area law. Physics Letters B, 686(2), 181-187.

\bibitem{KARA}
Karami, K., Sheykhi, A., Jamil, M., Azarmi, Z., and Soltanzadeh, M. M. (2011). Interacting entropy-corrected new agegraphic dark energy in Brans–Dicke cosmology. General Relativity and Gravitation, 43(1), 27-39.

\bibitem{ROVE}
Rovelli, C. (1996). Black hole entropy from loop quantum gravity. Physical Review Letters, 77(16), 3288.

\bibitem{GHOS}
Ghosh, A., and Mitra, P. (2005). Log correction to the black hole area law. Physical Review D, 71(2), 027502.

\bibitem{CAIY}
Cai, Y. F., Liu, J., and Li, H. (2010). Entropic cosmology: a unified model of inflation and late-time acceleration. Physics Letters B, 690(3), 213-219.

\end{thebibliography}
\end{document}